\documentclass[9pt]{article}
\usepackage{spconf,amsmath,graphics,graphicx}
\usepackage{xcolor}
\usepackage{enumitem}
\usepackage{amssymb}
\usepackage{amsmath}
\usepackage{siunitx}
\usepackage{empheq, etoolbox}
\patchcmd{\subequations}% <cmd>
  {\theparentequation\alph{equation}}% <search>
  {\theparentequation.\alph{equation}}% <replace>
  {}{}
\setlist{nosep, leftmargin=14pt}

\usepackage{mwe} % to get dummy images

\title{A deep residual learning implementation of Metamorphosis}
\name{}
\name{Matthis Maillard$^{\star}$ \qquad Anton François$^{\star \ddag}$ \qquad Joan Glaun\`es$^{\ddag}$ \qquad Isabelle Bloch$^{\star \dagger}$ \qquad Pietro Gori$^{\star}$}
\address{}
\address{$^{\star}$ LTCI, Télécom Paris, Institut Polytechnique de Paris, Paris, France \\ $^{\dagger}$ Sorbonne Universit\'e, CNRS, LIP6, Paris, France\\ $^\ddag$ MAP5, Universit\'e de Paris, France}
\begin{document}
\ninept
%https://www.overleaf.com/project/6135e7d10d30523763e1f94a
\maketitle
%https://www.overleaf.com/project/6135e7d10d30523763e1f94a
\begin{abstract}
 
In medical imaging, most of the image registration methods implicitly assume a one-to-one correspondence between the source and target images (\textit{i.e.,} diffeomorphism). However, this is not necessarily the case when dealing with pathological medical images (\textit{e.g.}, presence of a tumor, lesion, etc.).
To cope with this issue, the \textit{Metamorphosis} model has been proposed. It modifies both the shape and the 
appearance of an image to deal with the geometrical and topological differences. However, the high computational time and load have hampered its applications so far. Here, we propose a deep residual learning implementation of Metamorphosis that drastically reduces the computational time at inference. Furthermore, we also show that the proposed framework can easily integrate prior knowledge of the localization of topological changes (\textit{e.g.,} segmentation masks) that can act as spatial regularization to correctly disentangle appearance and shape changes. We test our method on the BraTS 2021 dataset, showing that it outperforms current state-of-the-art methods in the alignment of images with brain tumors.
%introduce a spatial regularization to correctly disentangle appearance and shape changes.
%control the localization of the intensity modification, ensuring it changes only the topology of the image and not its shape.

\end{abstract}


\section{Introduction}
\label{sec:intro}

%Computing a diffeomorphic mapping between two images is at the core 
Image registration has many applications in medical imaging, such as
%\PG{Image registration is widely used in} medical imaging, 
%it is used for 
%\PG{for example to perform}
statistical analysis, modality fusion, surgery planning and follow up. 
%post operation analysis.
Most of the methods assume that the two images have the same topology (\textit{i.e.,} same number of anatomical components) thus looking for a diffeomorphism~\cite{ashburner_fast_2007, beg_computing_2005}. However, in some studies one needs to align images characterized by a different number of anatomical components,
%some studies require images with a different number of components to be aligned,
like a healthy image and an image with a tumor~\cite{roux2019}.

One of the first strategies to deal with a difference in topology between source and target images was the cost function masking, where the lesion was excluded when computing the image similarity~\cite{brett_spatial_2001}. %The deformation in the masked area is determined by interpolating the transformations in the neighborhood voxels. 
However, when the lesion is too big, the deformation in the masked zone might be very distorted. To cope with that, geometric metamorphosis \cite{niethammer_geometric_2011} adds a specific deformation to the masked area, but it works only when the lesion/tumor is present in both source and target images.
%. Unfortunately, the authors assume 
%\Isa{but assuming that} the object is present in both the source and target image, \Isa{which is too strong an assumption for our purpose}.}  
Another strategy, when one wants to register two images with and without tumor, is to first model the growth of the tumor in the healthy image by using a biophysical model~\cite{gooya_glistr:_2012, scheufele_coupling_2019}, and then do the registration.
%The image is then registered to the brain image.
This strategy deals with the topological difference, since the images now contain the same number of components, but it requires a user initialization, extensive computations to estimate the model parameters, and they are specific to a particular kind of tumor.
%has a high computation time and it is not generic as it is applicable only in the presence of tumors and growth models are different for each tumor type
%solves the topology problem since the images contain the same number of components. However, this approach has a high computation time and it is not generic as it is applicable only in the presence of tumors and growth models are different for each tumor type. 
In~\cite{liu_low-rank_2015}, authors use a similar but inverse perspective where, instead of adding a tumor to a healthy image, they remove the pathological region by synthesizing a quasi-normal image via low-rank approaches. This approach can effectively recover tumor regions, but at the same time distort or blur the healthy regions. Furthermore, it is a statistical technique that needs lesions to be homogeneously (and randomly) distributed across the population~\cite{liu_low-rank_2015}, which is not the case for all kinds of lesions or tumors (\textit{e.g.,} glioblastoma). With a similar perspective, inpainting
techniques have also been proposed~\cite{sdika_nonrigid_2009}. However, they may produce inaccurate or unreal images when dealing
with large tumors or lesions.
Metamorphosis~\cite{trouve_metamorphoses_2005} is a more generic approach that offers theoretical guarantees to tackle both shape and appearance differences. It consists in repetitively applying (together) infinitesimal shape and appearance deformations to the source image. To the best of our knowledge, there is no clinical application of this method and a recent implementation~\cite{anton_metamorphoses} indicates that it is very time consuming.

Recently, motivated by faster computation time at inference, several deep learning approaches have been developed for image registration. Neural networks are trained on a database of image pairs and not only for a fixed pair of images~\cite{balakrishnan_voxelmorph_2019}. The training is thus very time consuming but this type of methods is much faster at inference time than previous methods that need to optimize a functional for every pair of images.
%once it is trained, the model can predict the deformation for any pair of images of the same type as the one of the input data. This makes it much faster at inference time than previous methods that need to optimize a functional for every pair of images. 
Similarly, deep learning can also be used to align images with different topology. In~\cite{bone_learning_2020}, authors proposed a Metamorphic (variational) Auto-Encoder (MAE) to modify both the geometry and the appearance of an image at the same time. %MAEs approximate the Metamorphosis model by separately estimating appearance and shape variations \Isa{j'enleverais cette phrase qui semble contradictoire avec le reste}. 
However, this method may result in poor disentanglement, where the intensity changes can actually modify the shape of the input image.
%However, the method was built for atlas construction, therefore not directly applicable to aligning two images. Moreover, the geometrical and intensity modifications were not totally disentangled as the intensity could modify the shape of the input image. 

Inspired by 
%Metamorphosis \cite{trouve_metamorphoses_2005} and 
\cite{amor_resnet-lddmm_2021}, we propose a ResNet-based \cite{he_deep_2016} implementation of Metamorphosis which overcomes all previous limitations. Our contributions are: 
\begin{itemize}
\item We use a residual deep network~\cite{he_deep_2016} to solve the system of differential equations of Metamorphosis (see Eq.~\ref{eq:Metamorphosis} next). 

\item Our method can be used: 1) to optimize an energy between two paired images, or 2) in a learning-based scheme where we use a training set of source images and one fixed target image.
    %where the target image is fixed but source image changes. The latter option allows for a very fast computation time.
    
\item We introduce a local regularization that consists in limiting the intensity changes to a pre-specified zone (segmentation mask). We show that this produces a better disentanglement between shape and appearance changes. %regularization prevents the appearance from modifying the shape of the input image.

\item We evaluate our method on a synthetic shape dataset and on the BraTS 2021 dataset~\cite{brats_2015}. 
    %to show that the method is generic and can be applied when a topology change occurs between two images.
\end{itemize}

\section{Method}
\label{sec:method}

\textbf{Mathematical formulation.}
 Let $\Omega \subset \mathbb{R}^d$ be a bounded domain, with $d\in\{2,3\}$. Let $V$ be a Reproducible Kernel Hilbert space (RKHS) with kernel $K$ of vector fields with support $\Omega$ and $T$ times continuously differentiable, where $T\in \mathbb{N}^*$. Let $I$ be a gray scale image defined as a square integrable %\Isa{où a-t-on besoin de l'intégrer ?} \MM{Le square integrable c'est pour utiliser la norme L2 sur l'image. Peut être ce n'est pas utile d'écrire cela dans un papier appliqué puisque toutes les images sont square integrable.}
 and differentiable function $I: \Omega \rightarrow \mathcal{R}$. The aim of Metamorphosis is to modify a source image $I$ so that it perfectly aligns with a target image $J$. The model joins diffeomorphic deformations with additive intensity changes.
 %Therefore, one needs to minimize a dissimilarity measure such as the sum of squared distance between $J$ and $I_T$, the result of the metamorphose. In addition, a regularization term ensures that the shape deformation is smooth and that the intensities are only modified as little as possible. 
 % Non, normalement c'est un exat probleme... il faut parler de ce point après.
Similarly to~\cite{trouve_metamorphoses_2005,anton_metamorphoses}, the evolution of the image $I$ at time $t\in [0, 1]$ is: 
\begin{equation}
  \partial_t I_t =  v_t \cdot I_t + \mu^2 z_t = - \langle \nabla I_t, v_t \rangle + \mu^2 z_t \quad \text{s.t. } I_0=I  
\end{equation}
where $v_t \cdot I_t$ implies that $I_t$ is deformed by an infinitesimal vector field $v_t \in V$ and $z_t: \Omega \rightarrow \mathbb{R}$ is the additive part corresponding to the infinitesimal intensity variation (called the residual image or momentum). The hyperparameter $\mu^2 \in \mathbb{R}^+$ balances the intensity and geometric changes.
As in~\cite{anton_metamorphoses}, we cast the metamorphic registration as an inexact matching problem minimizing the cost function:
\begin{equation}
E = \frac{1}{2}||I_1-J||^2_2 + \lambda[\int_0^1 ||v_t||_V^2 + ||\mu z_t||^2_2 dt] 
%\quad \lambda \geq 0
\end{equation}
where the first term is the classical $L_2$ data term (please note that other data terms could be used as well) and the second term, weighted by $\lambda$, is the regularization. It is composed of the total geometric and intensity energy of the deformation, respectively. As shown in~\cite{anton_metamorphoses}, the geodesic equations for Metamorphosis are:
\begin{subequations}
    \begin{empheq}[left=\empheqlbrace]{align}
    v_t &= -  K \ast (z_t\nabla I_t)\label{eq:v}\\
    \partial_t z_t &= - \nabla \cdot (z_tv_t)\label{eq:z}\\
    \partial_t I_t &= - \langle \nabla I_t, v_t \rangle + \mu^2 z_t\label{eq:I}
    \end{empheq}
    \label{eq:Metamorphosis}
\end{subequations}
with $||v_t||_V^2 = \langle z_t \nabla I_t, K \ast (z_t \nabla I_t )\rangle$, where $K$ is a Gaussian kernel.

\begin{figure}[ht]
    \centering
    \includegraphics[scale=0.3]{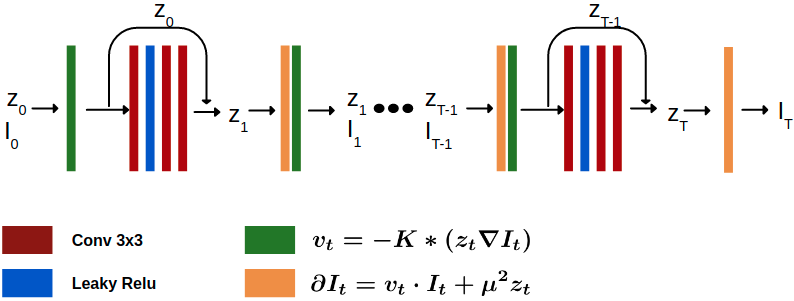}
    \caption{Architecture of the proposed method. }
    \label{fig:resnet-Metamorphosis}
    %\vspace{-0.5cm}
\end{figure}

\noindent From this system of equations, we can notice that $v_t$ is completely defined by $z_t$ and $I_t$, thus making $z_t$ the only unknown. The momentum $z_t$ has therefore a double role. It represents the additive intensity variation \textit{and} it is also the parameter of the deformation. This eases the computation but at the same time it makes the disentanglement between shape and intensity variations more difficult (see Fig.~\ref{fig:disentanglement}).
Inspired by \cite{amor_resnet-lddmm_2021,rousseau:hal-01796729}, we propose to use a residual neural network (ResNet) to find the solution of this system.
We take advantage of the similarity between ResNets and the numerical solutions of PDEs using Euler's method, given an initial value, to solve Eq.~\ref{eq:z}. Indeed, the numerical integration of Eq.~\ref{eq:z}, using discrete time steps $t$, is: %\Isa{using now discrete time steps, can be written as: }
\begin{equation}
z_{t+1} = z_t - \frac{1}{T} \nabla  \cdot (z_tv_t) \text{ for $t=i/(T-1)$ , $i \in 0,..,T-1$}
\end{equation}
where $T$ is the number of steps.
By replacing the divergence with a neural network, we obtain a ResNet:
\begin{equation}
    z_{t+1} = z_t + \frac{1}{T}f_{\theta_t}(z_t, I_t)
    \label{eq:f(z)}
\end{equation}
%ResNets can be modeled as:
%\begin{equation}
%x_{t+1} = x_t + f_{\theta_t}(x_t)
%\end{equation}
%\text{ for $t$ in $0,..,T-1$}$$
where $f_{\theta_t}$ is modeled as a convolutional layer followed by an activation function (Leaky ReLU) and two other convolutional layers.
%Isa{Mettre une note de bas de page ici pour souligner que la divergence (linéaire) est remplacée par un calcul non linéaire ? c'est d'ailleurs demandé par un des relecteurs}.
Compared with Eq. \ref{eq:z}, $v_t$ is replaced by $I_t$ because $v_t$ is a function of $I_t$ (and $z_t$).
%\PG{Need to say a word about this point. Rousseau uses other layers for each f. Maybe in the discussion.}. 
The network is built as a sequence of $T$ convolution blocks $f_{\theta_t}$.
At each time step $t$, $z_{t+1}$ is computed using Eq. \ref{eq:f(z)}. Subsequently, $v_{t+1}$ is calculated directly with Eq.~\ref{eq:v} and one determines $I_{t+1}$ by applying the geometric transformation induced by $v_t$ and adding the residuals $z_t$ as in Eq.~\ref{eq:I}. The architecture of the model is detailed in Fig.~\ref{fig:resnet-Metamorphosis}. 

\textbf{Optimization setting.}
The proposed method can be used in two different contexts: optimization or learning-based. 
%The first one is an optimization context where 
In the former, one wants to compute the transformation between two fixed images. In this case, the back-propagation algorithm is repetitively applied to minimize the following equation
%\ref{eq:optim} 
until convergence: 
\begin{equation}
    E_O(\theta,z_0) = \frac{1}{2}||I_T-J||^2_2 + \frac{\lambda}{T}\sum_{t=0}^{T-1}[||v_t||_V^2 + \mu^2 || z_t||^2_2]\label{eq:optim}
\end{equation}
%\Isa{pourquoi les deux sommes n'ont-elles pas le meme nombre de termes ? Aller jusqu'à $T$ pour les deux ? (et dans ce cas diviser par $T+1$ ? (voir aussi l'eq 7 ou la 2e somme commence à 1)}
The parameters to optimize are the weights of the neural network $\theta$ and the initial residuals $z_0$. %\Isa{donnés ? et donc dans ce cas les sommes devraient commencer à 1 ?}\MM{Non, on doit aussi regulariser $z_0$, car on le mets aussi à jour avec la descente de gradient. $z_0$ est initialisé à 0 au début.} 
We use Adam optimizer with a learning rate of $10^{-3}.$ %\PG{il faut dire que tu utilises un algo de minimization de gradient...et il faut dire par rapport à quoi tu calcules le gradient. Expliciter les paramètres }
%\PG{In this context, we have empirically checked that the ResNet approximates quite well the or is correlated ?? to the divergence in Eq.\ref{eq:z} (TO WRITE SOMETHING or Modeling Eq 3.b with a ResNet instead than directly computing a divergence brings to similar results but not exactly the same. Future research will dive more into that, trying to understand whether the use of a ResNet may give more generic and/or well adapted results.)}
In this context, we have empirically measured that the first layers are similar to the divergence in Eq.~\ref{eq:z}: their $L_1$ distance is small (between 0 and 1) and the Pearson's correlation coefficient is around 0.8. However, this is not the case for the last layers, where the $L_1$ distance is high (around 40) and the correlation is close to 0. Future research will dive more into that, trying to understand whether the use of a ResNet may give more generic and/or well adapted results.

%The second context is a learning context where one wants
In the learning-based setting, the goal is to compute the transformations between every image of a training dataset of $N$ images and a fixed target image. The optimized parameters are the same as in the previous context, the difference is that from one iteration to the other, the source image is not the same.  Therefore, one needs to minimize the following energy $E_L(\theta,z_0)$:
\begin{equation}
    \sum_{n=1}^N\left[\frac{1}{2}||I_T^n-J||^2_2 + \frac{\lambda}{T}[\sum_{t=0}^{T-1}||v^n_t||_V^2 + \mu^2 (||z_0||^2_2+\sum_{t=1}^{T-1}|| z^n_t||^2_2)]\right]
    \label{eq:learning}
\end{equation}
where $n$ is the index of image $I^n$ in the dataset. Please note that the learning setting is possible only because we use a neural network in Eq.\ref{eq:z}. The ResNet learns the optimal network parameters $\theta$ and initial residuals $z_0$ to align \textit{the entire} distribution of source images onto a (fixed) target image.
%Furthermore, we estimate the \text{same} $z_0$ for all images, which can be seen as the best initialization for the ResNet. \Isa{Justifier plus ? (cf relecteur 3).}
Once the network is trained, it can be used to register any image of the same type as those in the training set to the target image. Differently from the optimization setting, this allows for the registration of a high number of images in a very short time since the inference time of the ResNet is definitely lower compared to the optimization of Eq.~\ref{eq:optim} (less than one second vs one minute for images of size $200\times 200$). 
%Such scenario could occur when one needs to align a large set of images to a common template. 
This is particularly important, for instance, when one needs to normalize or align a large set of images to a common template for a statistical analysis.

\textbf{Local regularization.}
The main inconvenience with Metamorphosis is that it is hard to control the disentanglement between shape and appearance.
For instance, a trivial solution
%the the problem
 would be to set the overall geometrical deformation function to the identity (no geometrical change) and the overall appearance deformation map to $J-I_0$. In that case, the $L_2$ distance between the deformed image and $J$ would be 0 but it would not be a satisfactory result since homologous structures should be matched using only geometric deformations whereas appearance and topological changes ({\em i.e.}, new components) should be taken into account by the intensity modifications. The disentanglement can be controlled by tuning the hyper-parameters $\mu$ and $\lambda$. However, finding the right ones is a difficult task and they are different for each setting. If they are not correctly chosen, the appearance map could, for instance, modify the shape of the image, thus distorting the results and their interpretations.

To improve the control of the disentanglement, we introduce a local regularization which consists in limiting the region where the intensities can be modified (e.g. a tumor). To do so, we define such a region as $\alpha \subset \Omega$. We introduce a mask $m_0$ where $m_0(x) = 1$ if the pixel $x\in \alpha$ and 0 otherwise. %As the input image is modified geometrically $T$ times, the mask will also have to be geometrically modified $T$ times. 
During the deformation, the shape of the tumor will be modified, therefore, the mask must undergo the same deformation. Hence, we have $\partial_t m_t= v_t\cdot m_t$. The transformation of the image is then: $\partial_tI_t = v_t\cdot I_t +  \mu^2m_{t} z_t $. Using this equation and $||\sqrt{m_t}z_t||^2_2$ as regularization term for $z_t$, we obtain the same geodesic Equations \ref{eq:v} and \ref{eq:z}.

\section{Experiments}
\label{sec:experiments}
\begin{figure}[htbp]
\begin{minipage}[b]{\linewidth}
  \centering
\includegraphics[width=\textwidth]{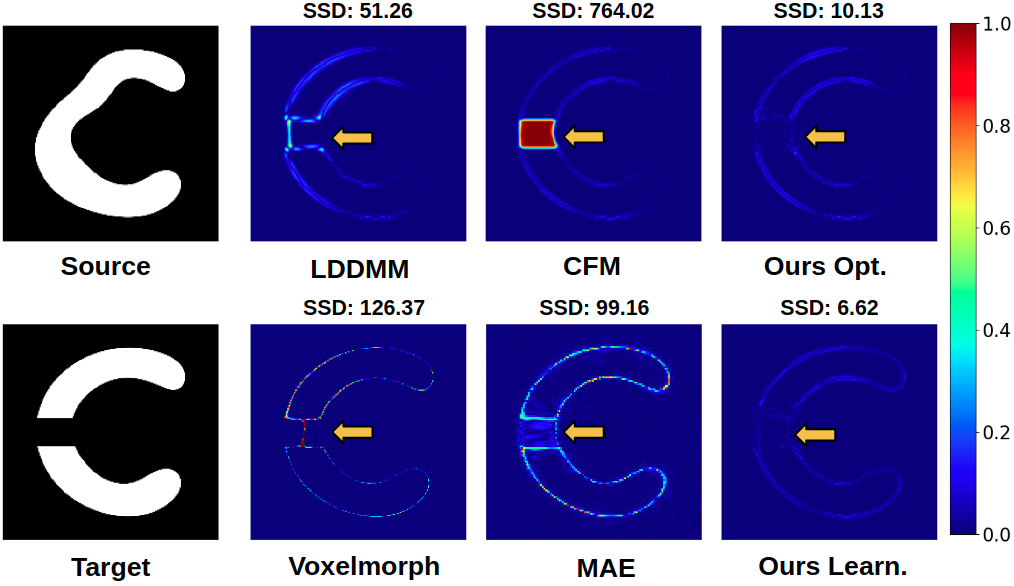}
\caption{Registration between Source and Target images for LDDMM~\cite{beg_computing_2005}, Cost Function Masking (CFM)~\cite{brett_spatial_2001}, ours in an optimization setting, Voxelmorph~\cite{balakrishnan_voxelmorph_2019}, Metamorphic Auto-Encoders (MAE)~\cite{bone_learning_2020} and ours in a learning context. Plots refer to the absolute value of the difference between the corresponding output and the target. SSD refers to the sum of squared differences. Top and bottom lines show the optimization and learning based methods, respectively. Arrows point to the topological difference between Source and Target image.}
%\vspace{-0.5cm}
\label{fig:comparison}
\end{minipage}
\end{figure}

\textbf{Datasets.}
We first evaluate the effectiveness of the method on a synthetic dataset of 2000 images of size $200\times200$. It is built from an image of a white ``C" on a black background. Each image of the dataset is generated by applying a random elastic deformation to the ``C" image. The target image associated with this dataset is a ``C" that has been cut in the middle, as shown in Fig.~\ref{fig:comparison}.

\begin{figure}[htbp]
\begin{minipage}[b]{\linewidth}
    \centering
    \includegraphics[width=\textwidth]{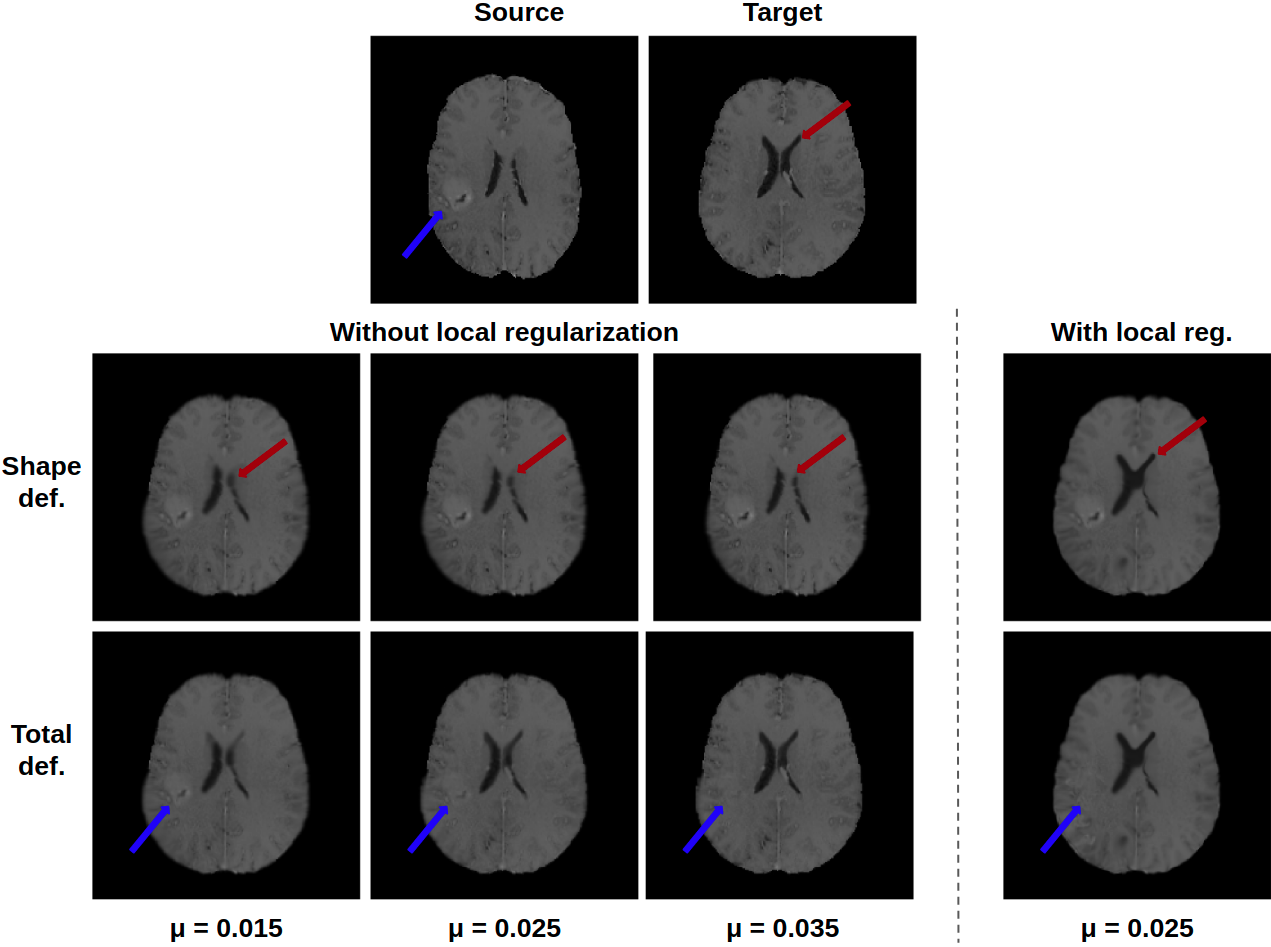}
    \caption{Comparison of the deformation between two T1ce images with and without local regularization. The Shape def. line only shows the geometrical deformation of the source, while Total def. also includes the estimated appearance changes. The tumor segmentation is used as mask $m_0$ for the local regularization.}
    \label{fig:disentanglement}
\end{minipage}
\end{figure}

The second dataset is BraTS 2021~\cite{brats_2015} comprising 4 MR modalities and the associated tumor segmentation image for 1251 brains with tumor. For each patient, we select the same slice of the T1 modality and crop it to obtain an image of size $208\times208$. We randomly pick 40 images from the dataset to form an evaluation set. The target image associated with this dataset is the linear MNI152 template \cite{mazziotta199588}. To show that our model also works with other modalities, we select a slice from the T1 contrast enhanced (T1ce) modality of two different patients. We extract one slice outside the tumor and the other one exhibiting part of the tumor (see Fig.~\ref{fig:disentanglement}). We use the segmentation of the tumor as mask for the local regularization.

\textbf{Results}
All experiments with our model have been computed using $T=20$ and $\lambda = 10^{-6}$.
The effectiveness of the method is first evaluated on the synthetic dataset. For comparison, we select optimization based methods: LDDMM \cite{beg_computing_2005} and cost function masking \cite{brett_spatial_2001}, and learning based methods: Voxelmorph \cite{balakrishnan_voxelmorph_2019} and Metamorphic auto-encoders (MAE) \cite{bone_learning_2020}. They are compared with our optimization and learning based method respectively. Voxelmorph and our learning model are trained with the ``C" dataset as source and the fixed target in Fig.~\ref{fig:comparison}. The source image in Fig.~\ref{fig:comparison} \textit{is not} included in the training set. MAE works the other way around, the source image is fixed and the target changes. Therefore, we create a dataset of images made of elastic deformations of the target image in Fig.~\ref{fig:comparison} and train MAE on this dataset. Similarly, the target image in Fig.~\ref{fig:comparison} \textit{is not} included in the training set. Thus, the three learning methods are comparable for the couple of images in Fig.~\ref{fig:comparison}. The visual and quantitative results show that, in both contexts, our method outperforms the others. As shown by the arrows, our model
can effectively deal with the topological change and 
correctly aligns the remaining part.
%the region where the topological change occurs. 

%We implemented the local regularization because the 
As already discussed, disentanglement between shape and appearance is hard to control. To illustrate this, we show the deformations between two T1ce images for several values of $\mu$ without regularization in Fig.~\ref{fig:disentanglement}. For $\mu = 0.035$, there is almost no geometrical deformation of the source, yet the final image is really close to the target. This indicates that $\mu$ is too big since all differences have been removed only by the appearance changes.
%because all the deformation has been dealt by the appearance deformation.
For the lower values of $\mu$: 0.015 and 0.025, the shape deformation is better but still unsatisfactory. As pointed out by the red arrow, the ventricles of the shape deformation image are not aligned with the target ones. However, the appearance deformation is also unsatisfactory because the tumor has not fully disappeared on both deformations. This shows that finding the optimal parameters to properly align two images \textit{without} regularization might be hard and time-consuming. By contrast, when using local regularization with the same $\mu$ values (here we only show $\mu= 0.025$), the model is able to properly align the two images. The shape deformation image, where we only deform the image without adding the intensity changes, shows that the ventricles have been correctly deformed. In the total deformation image, where we consider both geometric deformations and intensity changes, the tumor 
%is not visible anymore
has disappeared. Hence, with the proposed local regularization, shape and appearance are properly disentangled
%, thus demonstrating the usefulness of our approach towards this aim. Furthermore, 
and the method is more robust to the choice of $\mu$, since the results do not vary much when changing it.

\begin{figure}[htbp]
\begin{minipage}[b]{\linewidth}
    \centering
    \includegraphics[width=\textwidth]{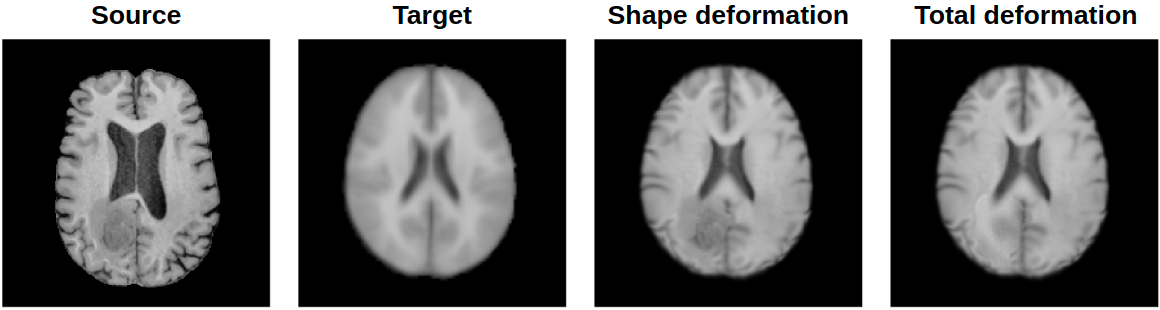}
    \includegraphics[width=\textwidth]{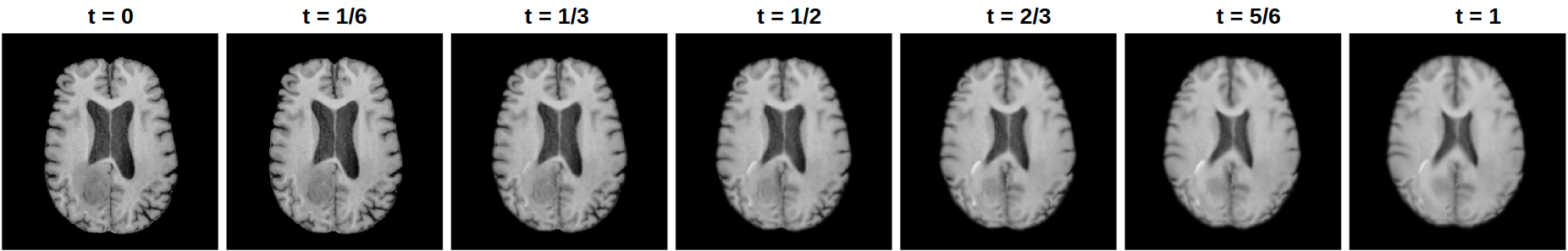}
    \caption{Deformation of a T1 image onto the MNI atlas. %The bottom line shows the evolution of the source image during the deformation. 
    }
    \label{fig:atlas}
    %\vspace{-0.5cm}
\end{minipage}
\end{figure}

Eventually, to show a potential application of the method in a clinical setting, we align a set of T1 images onto the MNI template. An example is shown in Fig.~\ref{fig:atlas}. To evaluate the quality of the alignment, we segment the ventricles of the deformed image and compute the Dice score with the reference segmentation of the ventricles in the MNI atlas (see Table \ref{tab:results}). As segmentation algorithm, we trained a U-Net model on 2D slices from the OASIS dataset~\cite{oasis}. Reference segmentations were obtained using FreeSurfer. The trained U-Net showed an average Dice score of 0.91 on the validation set. %\MM{Note that the output of the U-Net is not the mask $m_0$ for local regularization. It is only used to assess the quality of the registration.} %\PG{est-ce que tu mets des données avec des tumeurs dans ta base de training? est-ce que les images de Brats et OASIS sont pre-processed de la même façon?}\MM{Tu veux que je précise tout ça ? C'est pas sûr qu'on ait la place. J'utilise pas les images de Brats finalement parce qu'on obtenait les mêmes résultats. Même sur les images avec tumeurs la segmentation marche bien. J'applique le z-score avant le réseau pour OASIS et pour Brats}
%\PG{je pense qu'on peut dire alors que tu as utilisé OASIS et une partie de BRATS pour entrainer la méthode de segmentation. Si les résultats sont pareils ca ne change rien et ca evitera les questions embetants du genre "etes-vous sûrs que la méthode marche sur des images avec tumeurs?" après je pense que tu peux dire que tu as appliqué le même pre-processing aux données de OASIS e BRATS (même ceux utilisés pour le recalage).}
We compare our method, with and without local regularization ($\mu = 0.015$),  against cost function masking \cite{brett_spatial_2001} and Voxelmorph \cite{balakrishnan_voxelmorph_2019}. %Additionally, we evaluate our method without local regularization .
The presence of the tumor mainly modifies the shape of the ventricles (mass effect), but not their intensity. All methods should therefore correctly match the ventricles using only geometric transformations. For this reason, when using our methods, we compute the Dice score on the shape deformation images and not on the total deformation images.
%but rather than computing the scores on the total deformation, we do it on the shape deformation only. Since we showed before that the disentanglement between shape and appearance is not guaranteed, this allows to quantify the performance of the shape deformation without the regularization. Furthermore, we measure the time of each method. For the learning methods, only the inference times are shown. 
%We did not find any algorithm to segment ventricles in 2D images, therefore we trained a UNet model on 2D images. The model was trained on 2D slices of OASIS dataset~\cite{oasis} and was segmented by~\cite{hoopes2021} using FreeSurfer. We reached an average Dice score of 0.91 on the test set. 
%\Isa{en fait, a-t-on besoin d'expliquer ici comment les ventricules ont été segmentés ?}
%\PG{je pense qu'il faut dire 2 mots pour ne pas avori des questions...}
%\PG{RAPPEL: IL FAUT COMMENTER TABLE 1}
In Table \ref{tab:results}, we also show the SSD and inference time. Results clearly show that our method with local regularization better aligns the ventricles than the other methods and has the best SSD. This indicates that %modifying the intensities of the tumor
taking into account the topological changes during the deformation, rather than just masking them out or ignoring them, %helps to align the similar components
improves the alignment of homologous components between the two images.
%between the couple of images. 

\begin{table}[htbp]
    \centering
    \setlength{\tabcolsep}{0.7\tabcolsep}
    {\footnotesize
    \begin{tabular}{lccc}
    \hline
     Method    & SSD & Dice  & Inference Time (s)  \\
     \hline
     CFM & $167 \pm 68$ & $69.1\pm22.4$ & $55\pm1.2$ \\
     Voxelmorph & $187 \pm 73$ & $63.4\pm22.6$ & \boldmath$0.01\pm$8e-4 \\
     Ours w/o reg (learn.)& $220 \pm 20$&$65.4\pm20.7$&$0.15\pm0.003$\\
     Ours w/ reg (learn.)   & $161 \pm 16$&$69.8\pm15.2$&$0.16\pm0.009$\\
     Ours w/ reg (opt.)   & \boldmath$135 \pm 45$&\boldmath$71.9\pm19.5$&$59\pm1.2$\\
     \hline
    \end{tabular}
    }
    \caption{Quantitative evaluation for Cost Function Masking (CFM), Voxelmorph and our method with and without regularization. learn. (respectively opt.) indicates a learning context (respectively optimization context). Results were computed on a test set of 40 patients. Best results in bold.
    }
    %\vspace{-0.5cm}
    \label{tab:results}
\end{table}

\section{Conclusion}
This paper proposes a deep residual learning implementation of Metamorphosis and adds a local regularization to improve the shape and appearance disentanglement. Qualitative and quantitative results on BRATS dataset show the effectiveness of the method. Moreover, the model can be used in an optimization or learning context, the latter offering very fast computation time during inference.  

% Below is an example of how to insert images. Delete the ``\vspace'' line,
% uncomment the preceding line ``\centerline...'' and replace ``imageX.ps''
% with a suitable PostScript file name.
% -------------------------------------------------------------------------

%% To start a new column (but not a new page) and help balance the last-page
% % column length use \vfill\pagebreak.
% % -------------------------------------------------------------------------
% \vfill
% \pagebreak

% \section{Footnotes}
% \label{sec:foot}

% Use footnotes sparingly (or not at all!) and place them at the bottom of the
% column on the page on which they are referenced. Use Times 9-point type,
% single-spaced. To help your readers, avoid using footnotes altogether and
% include necessary peripheral observations in the text (within parentheses, if
% you prefer, as in this sentence).

% \section{Copyright forms}
% \label{sec:copyright}

% You must include your fully completed, signed IEEE copyright release form when
% you submit your paper. We {\bf must} have this form before your paper can be
% published in the proceedings.  The copyright form is available as a Word file,
% a PDF file, and an HTML file. You can also use the form sent with your author
% kit.

% \section{Referencing}
% \label{sec:ref}

% List and number all bibliographical references at the end of the
% paper.  The references can be numbered in alphabetic order or in order
% of appearance in the document.  When referring to them in the text,
% type the corresponding reference number in square brackets as shown at
% the end of this sentence \cite{C2}.

\textbf{Compliance with ethical standards}
This research was conducted retrospectively using human subject data made available in open access (OASIS and BRATS 2021 databases).

\textbf{Acknowledgments}
Authors report no conflict of interest. M. Maillard was supported by a grant of IMT, Fondation Mines-Télécom and Institut Carnot TSN.

% IEEE-ISBI supports the disclosure of financial support for the project
% as well as any financial and personal relationships of the author that
% could create even the appearance of bias in the published work. The
% authors must disclose any agency or individual that provided financial
% support for the work as well as any personal or financial or
% employment relationship between any author and the sources of
% financial support for the work.

% Other types of acknowledgements can also be listed in this section.

% Reporting on real or potential conflicts of interests, or the absence
% thereof, is required in the paper. Authors are responsible for
% correctness of the statements provided in the manuscript. Examples of
% appropriate statements include:
% \begin{itemize}
%   \item ``No funding was received for conducting this study. The
%     authors have no relevant financial or non-financial interests to
%     disclose.'' 
%   \item ``This work was supported by […] (Grant numbers) and
%     […]. Author X has served on advisory boards for Company Y.'' 
%   \item ``Author X is partially funded by Y. Author Z is a Founder and
%     Director for Company C.''
% \end{itemize}

\bibliographystyle{IEEEbib}
\bibliography{refs}

\end{document}